\documentclass[preprintnumbers,amsmath,12pt,amssymb,floatfix,superscriptaddress,nofootinbib]{article}

\def\mytitle#1{\setcounter{equation}{0}
\setcounter{footnote}{0}
\begin{center}\Large\textbf{#1}\end{center}
\vspace{0.25cm}}
\def\myname#1{\begin{center}{\large #1}\end{center}\vspace{-0.13cm}}
\def\myplace#1#2{\small\begin{center}\textit{#1}\\
\texttt{#2}\end{center}}

\def\myclassification#1{\small\noindent
Keywords :
       #1\vspace{0.5cm}}
\usepackage{amsmath}
\usepackage{color}
\usepackage{amsfonts}
\usepackage{amssymb}
\usepackage{breqn}
\usepackage{graphicx}
\usepackage{array}
\usepackage[utf8]{inputenc}
\usepackage{hyperref}
\usepackage{float}
\usepackage{subfig}
\usepackage{subcaption}
 \date{}
 
\begin{document}
\mytitle{Constraining Redshift Parametrization Models with Recentmost Data : Impacts on an Accretion Disc around Finslerian Kiselev Black Hole}

 \myname{Promila Biswas $^*$ \footnote{promilabiswas8@gmail.com$~~; ~~\text{Orchid} ~:~0009-0004-1431-3857 $},   Subhajit Pal $^*$\footnote{subhajitpal968@gmail.com$~~;~~\text{Orchid}~:~0009-0002-2431-1375$ }, Sukanya Dutta $^{**}$ \footnote{sduttasukanya@gmail.com$~~;~~\text{Orchid}~:~0009-0004-2989-309X$}, Ritabrata Biswas $^{**}$ \footnote{biswas.ritabrata@gmail.com$~~;~~\text{Orchid}~:~0000-0003-3086-892X$} and Farook Rahaman$^*$ \footnote{farookrahaman@gmail.com$~~;~~\text{Orchid}~:~0000-0003-0594-4783$}}
\myplace{$^*$Department of Mathematics, Jadavpur University, Kolkata-32, India\\ $^{**}$Department of Mathematics, The University of Burdwan, Burdwan-713104, India} {}


\begin{abstract}
We investigate the evolution of black hole mass within a cosmological background modeled by a Modified Chaplygin Gas (MCG) under various dark energy equation of state parametrizations, including Linear, Logarithmic, CPL, JBP models. The logarithmic mass ratio $\log_{10}[M(z)/M_0]$ is found to be highly sensitive to the redshift-dependent evolution of $\omega(z)$, with gentle slopes in Linear, Logarithmic and CPL models indicating quasi-static accretion and steep slopes in JBP corresponding to rapid late-time variations highlighting transient suppression or enhancement of accretion due to repulsive dark energy effects. Peaks, minima and amplitude offsets in the mass ratio reflect the 
dynamic interplay between horizon thermodynamics, the evolving pressure of the MCG and cosmic expansion, illustrating how the black hole mass growth is directly influenced by both the temporal evolution of dark energy and the effective gravitational potential of 
the surrounding cosmic fluid. Our results demonstrate that black hole accretion acts as a sensitive probe of the time-dependent cosmic pressure landscape and provides physical insights into the coupling between local strong gravity and global accelerated expansion.

\end{abstract}
\myclassification{Accretion, Dark energy, Black hole, Thermodynamics, Cosmological Parametrisation.}
\section{Introduction}


Dark energy (DE) is a theoretical form of energy that permeates all of space and accelerates the present day expansion of the universe. In cosmology, DE is often modelled as a perfect fluid with an equation of state (EoS) 
\begin{equation}\label{DEEoS}
p = \omega \rho~~~~,
\end{equation}
where $p$, $\omega$ and $\rho$ are the pressure, EoS parameter and energy density of the DE candidate respectively. For vanishing $\Lambda$, Friedmann equation  
$ \frac{\ddot{a}}{a} = - \frac{4 \pi G_N}{3} (\rho + 3p) $ for accelerated
expansion $(\ddot{a} >0)$ leads us to the condition $\rho+3p<0$, i.e, $\omega < -\frac{1}{3}$.

Einstein field equations with the cosmological constant $\Lambda$ is one way to model DE is written as, 
\begin{equation}
    G_{\mu \nu} + \Lambda g_{\mu \nu} = \frac{8 \pi G_N}{c^4} T_{\mu\nu}~~,
\end{equation}
where $G_{\mu\nu}, ~g_{\mu\nu}$ and $T_{\mu\nu}$ are Einstein tensor (curvature of space-time), metric tensor and energy momentum tensor respectively. Energy density associated with the cosmological constant is 
\begin{equation}
    \rho_{\Lambda} =  \frac{\Lambda{c^2}}{8 \pi G_N}~~~~,
\end{equation}
which remains constant over time, unlike the matter or radiation whose densities dilute as the universe expands. Cosmic coincidence problem, fine tuning etc were not justified by the cosmological constant and dynamical DE models were required to be introduced.

For a generalised DE model, first Friedmann equation turns
\begin{equation}
    H^2 = \frac{8 \pi G_N}{3} \left(\rho_b + \rho_{DE}+ \rho_r\right)+ \frac{\Lambda}{3}- \frac{\kappa}{a^2}~~~,
\end{equation}
where $\rho_b$, $\rho_{DE}$ and $\rho_{r}$ are respectively the density contribution by baryons, DE and radiation.
For expansion rate, we define
\begin{equation}
    E^2(z) \equiv \frac{H^2 (z) }{H_0^2} = \Omega_{b0} (1+z)^3 +\Omega_{DE0} F(z) + \Omega_{r0} (1+z)^4+ \Omega_{\kappa 0}(1+z)^2+\Omega_{0}~~~,
\end{equation}
where $\Omega_{b0},~\Omega_{DE0},~\Omega_{r0},~\Omega_{\kappa 0}~\text{and}~\Omega_{0}$ are present day dimensionless densities given as
\begin{equation}
    \Omega_i \equiv \frac{8 \pi G_N}{3 H^2} \rho_i~~~~~i=\text{rad, DE and DM}~~~.
\end{equation}
Due to curvature, the contribution is 
\begin{equation}
    \Omega_\kappa \equiv - \frac{\kappa}{H^2 a^2}
\end{equation}
and for cosmological constant, 
\begin{equation}
    \Omega_0 \equiv \frac{\Lambda}{3 H_0^2}~~~.
\end{equation}
$F(z)$ will be determined according to the model of DE chosen. 

As $E(z=0)=1$ at the present epoch,  we obtain 
\begin{equation}
    \Omega_{b0} + \Omega_{DE0}+ \Omega_{r0}+\Omega_{\kappa 0}+\Omega_{0}=1~~~~,
\end{equation}
where $ \Omega_{b0}$, $\Omega_{DE0}$, $\Omega_{r0}$ and $\Omega_{\kappa 0}$ are present time values of dimensionless densities. Depending upon this model, we will establish a tool to measure Hubble parameter at different redshifts. In section 3, such data points will be incorporated.

Black holes (BHs) are known to accrete matter and radiation. Babichev et al. (2004) \cite{babichev2005accretion} showed that BHs can accrete a perfect fluid with negative pressure. If DE behaves like a fluid with $\omega<-1$ (phantom energy), it could cause BHs to loose mass over time which is contrary to usual accretion. For $-1<\omega<-\frac{1}{3}$ (quintessence), BHs may still gain mass except a comparatively slow rate. Nayak and Singh (2011) studied the evolution of primordial BHs within the Brans-Dicke theory, incorporating phantom energy accretion \cite{nayak2011phantom}. Their findings indicated that radiation accretion increases the life time of these BHs, whereas phantom energy accretors accrete their evaporation. He et al \cite{He_2008} analysed the quasi normal modes of BHs absorbing DE. They found that the accretion of phantom energy could lead to growing modes in the perturbation tail, indicating potential instabilities. De Prolis, Jamil and Qadir \cite{de2010black} explored the effects of accreting viscous phantom energy onto Schwarzschild BHs. Their study revealed that bulk viscosity accelerates the mass loss of BHs compared to non viscous accretion, suggesting that the inclusion of viscosity in DE models can significantly alter BH dynamics. Jamil, Qadir and Rashid \cite{jamil2008charged} investigate charged BHs in a phantom environment. They find that the accretion of phantom DE onto a charged BH could eventually lead to the formation of a naked singularity.

Though DE does not directly couple to baryonic matter, interaction between DE and DM can still indirectly affect baryons by modifying the gravitational potential of DM halos which govern how baryons fall in and cool to form stars and galaxies. Since DM potential will define the gravitational environment for baryons, any evolution in DM density profiles due to DE-DM interaction can affect gas accretion rate, cooling efficiency of baryonic gas, star formation rates (SFRs), especially in early galaxies, etc. In other concepts, if DE transfers energy to DM, i.e., a positive coupling takes place, DM becomes effectively ``heavier", deepening the potential well and drawing in more baryons. Conversely, energy flow from DM to DE, if a negative coupling takes place, reduces the potential depth, making baryonic collapse less efficient. Change in halo potential affects supernova feedback, outflows and baryon retention. A shallower potential due to interaction might allow easier expansion of gas, suppressing star formation in low mass halos. Observationally, these can lead to altered baryonic Tully-Fisher relation \cite{McGaugh_2012} which links baryon mass to rotation speed. Impacts on stellar to halo mass relation and galaxy halo connection statistics are found. These models provide better fits to dwarf galaxy dynamics which show deviation from $\Lambda$CDM expectations.

Accretion occurs when a gravitational object like a BH interacts with a surrounding medium. For DE, this interaction depends on its energy-momentum flow, governed by GR. 

For a perfect fluid with
\begin{equation}
    T_{\mu \nu} = (\rho + p) u_{\mu} u_{\nu} + p g_{\mu \nu}~~~.
\end{equation}
If $\rho + p>0$, i.e., for normal or quintessence like DE, the BH mass. On the other hand, if $\rho + p<0$, i.e., for phantom DE, the BH looses mass due to a negative effective energy flux into the BH. If the chosen DE model is a scalar field, it can have spatial and temporal variation. In such models, DE has a non trivial energy flux, allowing it to accrete onto BHs. If DE dominates the cosmic fluid and a BH is embedded in such a background (e.g. Schwarzschild-deSitter on MeVittie space-time), accretion becomes theoretically feasible. 
\begin{equation}
    \rho_{DE} (z) = \rho_{DE,0} \times \exp{\left\lbrace{3 \int\limits_{0}^{z} \frac{1+ \omega(z^\prime)}{1+ z^\prime} dz^\prime}\right\rbrace}~~~.
\end{equation}
Current observations points towards time-varying DE models. Redshift parameterizations $\left({p(z)= \omega(z) \rho(z)}\right)$ provide a phenomenological, model-independent way to describe DE. Instead of assuming a specific field or potential (like quintessence), we explore the possible behaviours of $\omega(z)$ directly from data. Some parameterizations (like JBP or BA) are designed to behave well at both low and high redshift. Predictive power, i.e., to reveal the possible future of the universe is an addon utility of such models.

We enlist several redshift parametrization models in the table 2. In the second column, we recall their EoS-s. We analyze them for the data set noted in table 1 and constrain the free parameters and the best fits are noted in the 3rd column of table 2.

Studies regarding the accretion onto a BH which carries DE contamination in its metric results wind curves of radial velocity to end at different distances from the central compact object. This reduces the extremity of DE accretion onto Kerr metric which predicts radial inward speed wind to be equal to the speed of light at a finite distance. In this work, a DE contaminated black hole solution will be chosen and affect on its mass  will be followed. The DE candidate which affects from the BH metric itself and the candidates which will be modelled to accrete are dynamical. Variations of BH mass with respect to redshift will be studied. We will follow how the ratio of the mass to present day mass does change. 

In the next section (section 2), we will recall a BH solution which is sitting in the MCG filled universe. In section 3 and 4, we will collect and enlist $H(z)-z-\sigma(z)$ data points and constrain some redshift parameterization type DE models. In Section 5, accretion of previously constrained DE models onto the MCG contaminated BH will be studied. Finally, we will conclude this article in section 6.

\section{Finslerian Kiselev Black Hole Solution }
For vacuum, i.e., absence of matter or radiation leads to $T_{\mu\nu}=0$, i.e., $R_{\mu\nu}=\Lambda g_{\mu\nu}$ which governs the vacuum geometry in the presence of DE. The vacuum solution with $\Lambda>0$ is popularly known as the deSitter space-time. The deSitter space is maximally symmetric with constant positive curvature, described in static co-ordinates as $[c=1]$
\begin{equation}\label{ds}
ds^2 = f(r)\, dt^2 - \frac{dr^2}{f(r)} - r^2 \left(d\theta^2 + \sin^2\theta\, d\phi^2\right)~~~~,0\leq \theta \leq \pi~~,~~0\leq \theta \leq 2\pi~~\text{with}
\end{equation}
\begin{equation}
    f_{dS}=\left(1-\frac{\Lambda r^2}{3}\right)~~~~.
\end{equation}

Here, a cosmological horizon appears at $r=\sqrt{\frac{3}{\Lambda}}$. No observer can access region beyond this. This horizon is not due to matter, but purely due to space-time expansion, which is a pure effect of DE. 

A BH is a region in space-time where gravity is so intense that nothing, not even light can escape its inward pull. These compact objects are typically formed from the remnants of massive stars that collapse under their own gravity, BHs compress matter into an incredibly small region with infinite density, known as a singularity. Although BHs themselves are invisible, their presence is detected through the gravitational effects on nearby stars and gas\cite{Ghez2008}, as well as by observing high energy radiation emitted by matter being pulled in. 

Now, if a matter mass $M$ is thought to be placed in deSitter space, we reach at the Schwarzschild-deSitter(SdS) metric given by equation (\ref{ds}) with the lapse function
\begin{equation}\label{SdS}
f_{SdS}(r)=1-\frac{2G_NM}{r}-\frac{\Lambda r^2}{3}~~~~.
\end{equation}
The second term in equation (\ref{SdS}) is gravitational attraction and the last one is the repulsion due to DE. Here, we find two horizons, viz., BH event horizon($r_{bh}$) and cosmological horizon ($r_c$). At small radial distance, gravity dominates and at large radial distance, DE dominates. The region between $r_{bh}$ and $r_c$ is static and observable. However, two horizons lead to two surface gravities, i.e., two different temperatures and different thermodynamic properties are evolved.

If now, we opt Reissner-Nordstrom-deSitter (RNdS) BH, lapse function of this spherically symmetric, charged BH comprising a cosmological constant looks like
\begin{equation}
\text{where}~~    f_{RNdS}(r)=1-\frac{2G_NM}{r}+\frac{G_NQ^2}{4\pi\epsilon_0 r^2}-\frac{\Lambda r^2}{3}~~~~,
\end{equation}
where $Q$ is electric charge and $\epsilon_0$ is the electric constant. A vanishing lapse function gives upto three real positive roots, corresponding to inner(Cauchy) horizon $r_{-}$, outer Cauchy horizon $r_{+}$ and cosmological horizon $r_c$. A condition $r_{-}<r_{+}<r_c$ occurs for specific ranges of $M$, $Q$ and $\Lambda$. High charge on small mass lead to a naked singularity\cite{Joshi1993}.

Moving one step ahead, Kerr deSitter, i.e., rotating BH in deSitter space can be written. Besides three horizons mentioned earlier, ergospheres can be obtained. These ergospheres' sizes and shapes can get affected by $\Lambda$ as well\cite{Akcay2011}.

Including quintessence in Schwarzschild or Reissner-Nordstrom solutions, Kiselev BH is proposed with the lapse function ($G_N=1$)
\begin{equation}
f_{Kiselev}(r)=1-\frac{2M}{r}-\frac{c}{r^{3\omega+1}}~~~~,
\end{equation}
when $\omega=-1$, we are back to the deSitter background (\ref{SdS}). Using Newmann-Janis algorithm \cite{newman_janis_1965}, rotating Kiselev BHs are calculated by Ghosh et al\cite{Ghosh2015, Ghosh2023}.

On a differentiable manifold $M$, we consider a Finsler metric having the form $F=F(x, ~y)$ and being the function of $(x^i,~ y^j)$ in the tangential space of $M$. The Finslerian geodesic with respect to $F$ is characterised by 
\begin{equation}\label{Geodesic}
    \frac{d^2x^v}{d\tau^2}+\frac{1}{2}g^{v\omega}\left(\frac{\partial^2F^2}{\partial x^k \partial y^{\omega}}y^k-\frac{\partial F^2}{\partial x^{\omega}}\right)=0~~~~.
\end{equation}
The metric coefficient $g_{v\omega}$ can be written as 
$$g_{v\omega}=\frac{\partial ^2}{\partial x^k \partial y^{\omega}}\left(\frac{F^2}{2}\right)~~~~~.$$
We write the Ricci scalar of Finsler structure as 
$$Ric=R_\mu^\mu=\frac{1}{F^2}\left(2\frac{\partial G^\mu}{\partial x^{\mu}}-y^\nu\frac{\partial^2 G^\mu}{\partial x^\nu \partial y^\mu}+2G^{\nu}\frac{\partial^2G^\mu}{\partial y^{\nu}\partial y^{\mu}}-\frac{\partial G^{\mu}}{\partial y^\nu}\frac{\partial G^{\nu}}{\partial y^\mu}\right)~~,$$
and Ricci tensor as
$$Ric_{v\omega}=\frac{\partial^2}{\partial y^v \partial y^w}\left(\frac{1}{2}F^2Ric\right)~~~.$$

The angular coordinate was taken into account into account in the following ansatz as 
\begin{equation}\label{Finsler_structure}
    \bar{F}^2(\theta,~\phi,~y^{\theta}, ~y^{\phi})=\frac{1}{r^2}\left\{e^\nu y^t y^t-e^\mu y^ry^r-F^2\right\}~~~,
\end{equation}
where $\nu=\nu(r)$ and $\mu=\mu(r)$ and $\frac{\partial \bar{F}}{\partial y^t}=\frac{\partial \bar{F}}{\partial y^r}=0$. Therefore finsler metric coefficients can be obtained as follows
$$g_{v\omega}=diag\left(e^\nu,~ -e^\mu, -r^2\bar{g}_{ij}\right)~~\text{and} ~~g^{v\omega}=diag\left(e^{-\nu},~ -e^{-\mu}, -r^{-2}\bar{g}_{ij}\right)~,$$
By inserting the Finsler structure\eqref{Finsler_structure} into the geodesic \eqref{Geodesic}, we have 
\begin{equation}
   G^t=\frac{1}{2}v'y^ty^r,~~~ G^r=\frac{\nu'}{4}e^{\nu-\mu}y^ty^t+\frac{\mu'}{4}y^ry^r-\frac{r}{2}e^{-\mu}\bar{F}^2,~~~  G^{v}=\frac{1}{r}y^vy^r+\bar{G}^{v}~.~~~(v=\theta,~\phi)
\end{equation}
$\bar{G}^v$ is the geodesic spray coefficient derived by $\bar{F}$. Combining, we obtain 
$$Ric F^2=\left[\frac{1}{2}\left(v''+(v')^2\right)e^{\nu-\mu}-++++\dots\right]~~~,$$
In the Finsler structure, the scalar curvature turns into 
$$S=g^{v\omega}Ric_{v\omega}=\left(v''+(v')^2\right)e^{-\mu}\dots~~~.$$
In the context of the Rastall theory of gravity, we are searching for the general nonvacuum spherically symmetric static charged BH solutions in the section. The Rastall field equations can thus be expanded as $$G_{v\omega}+\kappa \lambda g_{v\omega}S=\kappa T_{v\omega}~~,$$
where $\lambda$ is the Rastall parameter.

When we consider a static, spherically symmetric BH immersed in a field made up of dust (d), radiation(r), quintessence(q), cosmological constant ($\Lambda$) or even any combination of these nonvanishing Maxwell tensor components as $E_j^i=\frac{Q^2}{\kappa r^4}diag\left(-1,~-1,~1,~1\right)$. Therefore, we can consider the nonzero entropy momentum tensor components as follows
$$T_t^t=T_r^r=\rho_s+\frac{Q^2}{\kappa r^4}~~, $$
$$T_{\theta}^{\theta}=T_{\phi}^{\phi}=-\frac{1}{2}(1+3\omega_s)\rho_s-\frac{Q^2}{\kappa r^4}~~,$$
where $\rho_s$ and $\omega_s$ are the density and equation of state parameter respectively. Symmetry of the problem helps us to consider
$$-\nu=\mu=-ln( _Ff_s(r))~~.$$
Therefore
$$G_t^t=G_r^r=-\frac{1}{r^2}\left(r~_Ff'_s(r)+ _Ff_s(r)-\bar{Ric}\right)~~~~,$$
$$G_\theta^\theta=G_\phi^\phi=-\frac{1}{r^2}\left(\frac{1}{2}r^2~_Ff''_s(r)+r~_Ff'_s(r)\right)~~~~\text{and}$$
$$S=\frac{1}{r^2}\left(r^2~_Ff_s''(r)+4r~_Ff'_s(r)+2~_Ff_s(r)-2\bar{Ric}\right)~~.$$
The $H_t^t=\kappa T_t^t~,~H_r^r=\kappa T_r^r~~\text{and}~~H_\theta^\theta=\kappa T_\theta^\theta$ components of Rastall field equations of the following differential equations result in 
$$-\frac{1}{r^2}\left(r~_Ff'_s(r)+ _Ff_s(r)-\bar{Ric}\right)+\frac{\kappa \lambda}{r^2}\left(r^2f''_s(r)+4r~_Ff_s'(r)+2~_Ff_s(r)-2\bar{Ric}\right)=\kappa \rho_s+\frac{Q^2}{r^4}~~~~,$$
\begin{equation}\label{set_of_diffequn}
    -\frac{1}{r^2}\left(\frac{1}{2}r^2~_Ff''_s(r)+r~_Ff'_s(r)\right)+\frac{\kappa \lambda}{r^2}\left(r^2~_Ff''_s(r)+4r~_Ff'_s(r)+2~_Ff_s(r)-2\bar{Ric}\right)=-\frac{1}{2}(1+3\omega_s)\kappa \rho_s -\frac{Q^2}{r^4}~~.
\end{equation}
Now, one can derive the following general solution for the metric function by solving the set of differential equations \eqref{set_of_diffequn},
$$ _Ff_s(r)=\eta -\frac{2M}{r}+\frac{Q^2}{r^2}-\frac{N_s}{r^{\frac{1+3\omega_s-6\kappa\lambda(1+\omega_s)}{1-3\kappa \lambda(1+\omega_s)}}}~~~~,$$
with energy density as
$$\rho_s=\frac{-3\omega_sN_s}{\kappa r^{\frac{3(1+\omega_s)-12\kappa\lambda(1+\omega_s)}{1-3\kappa \lambda(1+\omega_s)}}}~~~~,$$
where the quantities with $F$ denote the quantities of FKBH, $\bar{Ric} =\eta=constant$
$$\omega_s=-\frac{(1-4\kappa \lambda)(k\lambda(1+\omega_s)-\omega_S)}{(1-3\kappa \lambda(1+\omega_s))^2}~~~~.$$
This BH looks like a charged Kiselev like BH with a deficit solid angle.

\section{Differential Ages Method : Collection of Hubble parameter vs Redshift data}
From the definition of Hubble parameter at redshift $z$, i.e., $H(z)=\frac{\dot{a}}{a}$ and the relation between the redshift $z$ and the scale factor $a=\frac{1}{1+z}$, differentiating with respect to cosmic time,
$$\frac{dz}{dt}=\frac{d}{dt}(1+z)=-\frac{\dot{a}}{a^2}=-\frac{H(z)}{a}=-H(z)(1+z)\implies H(z)=-\frac{1}{1+z}\frac{dz}{dt}~~~~.$$
We are with fundamental equation of the different ages method. Since we cannot measure the exact derivative $\frac{dz}{dt}$ directly from observations, we use nearly data points to approximate it using finite differences. This gives $$\frac{dz}{dt} \approx \frac{\Delta z}{\Delta t}~~~~.$$

Ages are determined by stellar population synthesis models from galaxy spectra. For this, those galaxies are chosen which are massive, passively evolving (i.e., no significant star formation) and existing over a narrow redshift range. From this, classes of two galaxies are opted with redshift and age $(z,~t(z))$ and $(z+\Delta z,~t(z + \Delta z))$ (say) respectively. Then
$$H(z) \approx -\frac{1}{1+z}.\frac{(z+\Delta z)-z}{t(z+\Delta z)-t(z)}~~~.$$ Assuming $\Delta z = z_2 -z_1 \ll 1.$

Since Hubble parameter estimate is a ratio, uncertainties $\Delta t$ dominate the error bars
$$H(z) = -\frac{1}{1+z} \frac{\Delta z}{\Delta t} ~~~\implies \delta H(z) = \frac{1}{1+z} \frac{\Delta z}{\Delta t^2} \delta(\Delta t)~~~. $$

This is a direct measurement of $H(z)$ unlike integral methods (SNe, BAO) that gives distances. No assumption of spatial flatness, DE model, etc. This differentiates well between different DE models through slope of $H(z)$. 

To check the uncertainty, we denote uncertainties in redshift and age difference respectively as $\sigma_{\Delta z}$ and $\sigma_{\Delta t}$ and obtain,
$$\sigma_H^2 = \left( \frac{\partial H}{\partial \Delta z} \right)^2 + \left( \frac{\partial H}{\partial \Delta t} \right)^2 \sigma^2 _{\Delta t}~~~.$$
Hence, the partial derivatives are calculated as 
$$\frac{\partial H}{\partial \Delta z} = -\frac{1}{1+z} \frac{1}{\Delta z}~~~\text{and}$$
$$\frac{\partial H}{\partial \Delta z} = \frac{1}{1+z} \frac{\Delta z}{\Delta t^2}~~~,$$
leaving the total uncertainty as,
$$\sigma^2_H = \left(\frac{1}{1+z} \frac{1}{\Delta t} \right)^2 \sigma^2 _{\Delta z} + \left(\frac{1}{1+z} \frac{\Delta z}{\Delta t^2} \right)^2 \sigma^2 _{\Delta t}~~~.$$
Using differential ages method, different Hubble parameter data are obtained which are enlisted in table 1.

\begin{table}[H]
\caption{Hubble parameter $H(z)$ with redshift and errors $\sigma_H$ from Differential Ages method}
 \begin{minipage}{.5\linewidth}
\label{DA method}
\centering
	\begin{tabular}{ ||c|c|c|c|| }		
	\hline
		Sl No. & z & $H(z)$ with $\sigma_H(z)$ & Ref \\ 
		\hline
		1 & 0 & $69.01 \pm 1.30$ & \cite{Farooq2017} \\
		\hline
		2 & 0.07 & $69 \pm 19.6$ & \cite{Zhang2014RAA} \\ 
		\hline
        3 & 0.07 & $70.4 \pm 20$ & \cite{Zhang_2016} \\
        \hline
		4 & 0.09 & $69 \pm 12$ & \cite{Simon:2004tf}  \\
		\hline
        5 & 0.09 & $70.4 \pm 12.2$ & \cite{Simon:2004tf}\\
        \hline
		6 & 0.1 & $69 \pm 12$ & \cite{2010JCAP}  \\ 
		\hline
        7 & 0.1 & $70.4 \pm 12.2$ & \cite{Zhang_2016}\\
        \hline
		8 & 0.12 & $68.6 \pm 26.2$ & \cite{Zhang2014RAA} \\
        \hline
        9 & 0.12 & $70 \pm 26.7$ & \cite{Farooq2017}\\
		\hline
		10 & 0.17 & $83 \pm 8$ & \cite{2010JCAP} \\
		\hline
        11 & 0.17 & $84.7 \pm 8.2$ & \cite{Zhang_2016}\\
        \hline
		12 & 0.179 & $75 \pm 4$ & \cite{2012JCAP} \\ 
		\hline
        13 & 0.179 & $76.5 \pm 4$ & \cite{Moresco2012} \\ 
		\hline
        14 & 0.199 & $76.5 \pm 5.1$ & \cite{Moresco2012}\\
        \hline
		15 & 0.1993 & $75 \pm 5$ &  \cite{2012JCAP, Gomez-Valent:2018hwc}\\
		\hline
		16 & 0.2 & $72.9 \pm 29.6$ & \cite{Zhang2014RAA}  \\
		\hline
		17 & 0.24 & $79.7 \pm 2.7$ & \cite{Gaztanaga:2008xz}\\
        \hline
		  18 & 0.27 & $70 \pm 14$ &
        \cite{Simon:2004tf}\\
        \hline
        19 & 0.27 & $78.6 \pm 14.3$ & \cite{2010JCAP,Simon:2004tf}  \\
		\hline
        20 & 0.28 & $88.8 \pm 36.3$ & \cite{Zhang_2016} \\
        \hline
        21 & 0.28 & $88.8 \pm 36.6$ & \cite{Zhang2014RAA} \\
		\hline
        22 & 0.28 & $90.6 \pm 37.3$ & \cite{Farooq2017} \\
        \hline
		23 & 0.30 & $31.7 \pm 6.22$ & \cite{Zhang2014RAA} \\
		\hline
		24 & 0.35 & $82.7 \pm 8.4$ & \cite{Chuang_2013} \\
		\hline
		25 & 0.3519 & $83 \pm 14$ & \cite{Moresco_2015} \\
		\hline
        26 & 0.352 & $84.7 \pm 14.3$ & \cite{Moresco2012}\\
        \hline
		27 & 0.38 & $81.5 \pm 1.9$ & \cite{Alam:2016hwk}\\
		\hline
		28 & 0.3802 & $83 \pm 13.5$ & \cite{Moresco_2016} \\
		\hline
        29 & 0.4 & $87 \pm 17.4$ & \cite{Simon:2004tf}\\
        \hline
		30 & 0.4 & $95 \pm  17$ & \cite{Simon:2004tf} \\
		\hline
		\end{tabular}
\end{minipage}
\begin{minipage}{.5\linewidth}
	\centering
	\begin{tabular}{ ||c|c|c|c|| }
 		\hline
		Sl No. & z & $H(z)$ with $\sigma_H(z)$ & Ref. \\ [0.5ex] 
		\hline
		31 & 0.4004 & $77 \pm 10.2$ & \cite{Moresco_2016}  \\
		\hline
        32 & 0.4004 & $778.6\pm 10.4$ & \cite{Moresco_2016}  \\
		\hline
		33 & 0.4247 & $87.1 \pm 11.2$ & \cite{Moresco_2016} \\
		\hline
        34 & 0.4247 & $88.9 \pm 11.4$ & \cite{Moresco_2016} \\
		\hline
		35 & 0.43 & $86.5 \pm 3.7$ & \cite{Gaztanaga:2008xz}\\
		\hline
        36 & 0.43 & $88.3 \pm 3.8$ & \cite{Farooq2017}\\
		\hline
        37 & 0.44 & $82.6 \pm 7.8$ & \cite{Blake_2012} \\
		\hline
        38 & 0.44 & $84.3 \pm 7.9$ & \cite{Blake_2012} \\
		\hline
		39 & 0.4497 & $92.8 \pm 12.9$ & \cite{Moresco_2016} \\
		\hline
        40 & 0.4497 & $94.7 \pm 13.1$ & \cite{Moresco_2016} \\
		\hline
		  41 & 0.47 & $89 \pm 34$ &  \cite{Ratsimbazafy_2017}\\
        \hline
        42 & 0.47 & $89 \pm 49.6$ &  \cite{Gomez-Valent:2018hwc, Ratsimbazafy_2017}\\
		\hline
        43 & 0.47 & $90.8 \pm 50.6$ &  \cite{Ratsimbazafy_2017}\\
		\hline
		44 & 0.4783 & $80.9 \pm 9$ & \cite{Moresco_2016}\\
		\hline
        45 & 0.4783 & $82.5 \pm 9.2$ & \cite{Ratsimbazafy_2017}\\
		\hline
		46 & 0.48 & $97 \pm 9$ & \cite{2010JCAP}  \\
		\hline
        47 & 0.48 & $99 \pm 63.2$ & \cite{Stern_2010} \\
        \hline
		48 & 0.51 & $90.4 \pm 1.9$ &  \cite{Alam:2016hwk}\\
		\hline
        49 & 0.57 & $92.9 \pm 7.855$ & \cite{Anderson_2014}  \\
		\hline
		50 & 0.57 & $96.8 \pm 3.4$ & \cite{Anderson_2014}  \\
		\hline
		51 & 0.593 & $104 \pm 13$  & \cite{2012JCAP}\\
		\hline 
        52 & 0.593 & $106.1 \pm 13.3$  & \cite{Moresco_2016}\\
		\hline
		53 & 0.6 & $87.9 \pm 6.1$ & \cite{Blake_2012}\\
		\hline
        54 & 0.6 & $89.7 \pm 6.2$ & \cite{Zhang_2016}\\
		\hline
		55 & 0.61 & $97.3 \pm 2.1$ & \cite{Alam:2016hwk}\\
		\hline
        56 & 0.6797 & $92 \pm 8$ & \cite{Moresco_2016}\\
		\hline
        57 & 0.68 & $92 \pm 8$ & \cite{2012JCAP}\\
		\hline 
        58 & 0.68 & $93.9 \pm 8.1$ & \cite{Moresco_2016}\\
		\hline
		59 & 0.73 & $97.3 \pm 7$ & \cite{Blake_2012}\\
		\hline
        60 & 0.73 & $99.3 \pm 7.1$ & \cite{Blake_2012}\\
		\hline
    \end{tabular}
\end{minipage}	
\end{table}

\begin{table}[h!]
\begin{minipage}{.5\linewidth}
\centering
	\begin{tabular}{ ||c|c|c|c|| }		
	\hline
		Sl No. & z & $H(z)$ with $\sigma_H(z)$ & Ref \\ 
		\hline
        61 & 0.781 & $107.1 \pm 12.2$ & \cite{Moresco_2016}\\
		\hline
		62 & 0.782 & $105 \pm 12$ & \cite{2012JCAP}\\
		\hline
        63 & 0.875 & $127.6 \pm 17.3$ & \cite{Moresco_2016}\\
		\hline
		64 & 0.8754 & $125 \pm 17$ & \cite{2012JCAP}\\
		\hline
		65 & 0.88 &  $90 \pm 40$ & \cite{2010JCAP} \\
		\hline
        66 & 0.88 &  $91.8 \pm 40.8$ & \cite{Stern_2010} \\
		\hline
        67 & 0.88 & $117 \pm 23.4$ & \cite{Simon:2004tf}\\
        \hline
        68 & 0.9 & $69 \pm 12$ & \cite{Simon:2004tf} \\
		\hline
		69 & 0.9 & $117 \pm 23$ & \cite{2010JCAP} \\
		\hline
        70 & 0.9 & $119.4 \pm 23.4$ & \cite{Simon:2004tf} \\
		\hline
		71 & 1.037 & $154 \pm 20$ & \cite{2012JCAP}\\
		\hline
        72 & 1.037 & $157.2 \pm 20.4$ & \cite{Moresco_2016}\\
		\hline
		73 & 1.3 & $168 \pm 17$ & \cite{2010JCAP} \\
		\hline
		\end{tabular}
\end{minipage}
\begin{minipage}{.5\linewidth}
	\centering
	\begin{tabular}{ ||c|c|c|c|| }
 		\hline
		Sl No. & z & $H(z)$ with $\sigma_H(z)$ & Ref. \\ [0.5ex] 
		\hline
        74 & 1.363 & $160 \pm 33.6$ & \cite{Moresco_2015}\\
		\hline
        75 & 1.363 & $163.3 \pm 34.3$ & \cite{Moresco_2015}\\
		\hline
		76 & 1.43 & $177 \pm 18$ & \cite{2010JCAP} \\
		\hline
		77 & 1.53 & $140 \pm 14$ & \cite{2010JCAP, Simon:2004tf} \\
		\hline
        78 & 1.53 & $142.9 \pm 14.2$ & \cite{Simon:2004tf} \\ 
        \hline
        79 & 1.75 & $202 \pm 40$ & \cite{2010JCAP} \\
		\hline
        80 & 1.75 & $206.1 \pm 40.8$ & \cite{Simon:2004tf} \\
		\hline
		81 & 1.965 & $186.5 \pm 50.4$ & \cite{Moresco_2015}\\
		\hline
        82 & 1.965 & $190.3 \pm 51.4$ & \cite{Moresco_2015}\\
		\hline
		83 & 2.3 & $224.0 \pm 8.0$ & \cite{Busca_2013}\\
		\hline
		84 & 2.34 & $222 \pm 7$ & \cite{Delubac_2015}\\
		\hline
		85 & 2.36 & $226 \pm 8$ & \cite{Font_Ribera_2014}\\
        \hline
    \end{tabular}
\end{minipage}	
\end{table}
\section{Constraining Free Parameters of Redshift Parametrization Models using table 1}

The chi squared value quantifies the total squared difference between the observed and predicted expansion rates, scaled by observational uncertainty
\begin{equation}
\chi^2_{OHD}(H_0,~\tilde{\theta})= \sum_{i=1}^{N} \frac{\left\lbrace{H_{th} \left({z_i, \tilde{\theta}}\right)-H_{obs} (z_i)}\right\rbrace^2 }{\sigma_zi^2}~~~,
\end{equation}
where $H_{obs}(z_i)$ is the observed value of the Hubble parameter at redshift $(z_i)$ and  $H_{th} \left({z_i, \tilde{\theta}}\right)$ is the theoretical value from a model depending on parameters $\tilde{\theta}$ (eg. $H_0$, $\Omega_m$, $\omega_0^{DE}$ and $\omega_i^{DE'}$s etc.) $\sigma_{z,i}^2$ denotes the $1\sigma$ uncertainty  in the observed $H(z_i)$. $N$ is the total count of data points.

Inclusion of the Baryon Acoustic Oscillation (BAO) peak parameter in a cosmological data analysis - especially in combination with $H(z)$ measurements - enhances the constraining power on cosmological parameters. BAO is a ``Standard ruler'' imprinted in the distribution of galaxies, originally from sound wave in the early universe. A commonly used BAO peak parameter is the volume averaged distance, also called the dilation scale
$$D_v(z) = \left[ (1+z)^2 D_A^2 (z) \frac{cz}{H(z)} \right]^{\frac{1}{3}}~~~~~,$$
where $D_A(z)$ is the angular diameter distance, $c$ is the speed of light. Another common observable is 
$$A(z) = \frac{D_v(z) \sqrt{\Omega_m H_0^2}}{z}~~~~~~~ \cite{Eisenstein_2005}~~~.$$

This is called BAO acoustic parameter, usually measured at specific redshifts, 

If we include the BAO acoustic parameter measurement,
\begin{equation}
    \chi_{BAO}^2 = \frac{ \left[{\mathcal{A}}_{obs}(z)-{\mathcal{A}}_{th}(z ; \tilde{\theta}) \right]^2}{(\sigma_A)^2}~~~,
\end{equation}
\begin{equation}
    \chi_{BAO}^2 = \frac{({\mathcal{A}}_{obs}(z)-0.469)^2}{(0.017)^2}~~~,
\end{equation}
where we have used the measured value of ${\mathcal{A}}_{th}(z ; \tilde{\theta})$ and $\sigma_A$ are $0.469$ and $0.017$ respectively as obtained by \cite{Eisenstein_2005}.

The Cosmic Microwave Background encodes the state of the early universe - particularly the geometry and expansion at the surface of last scattering (redshift $z \approx 1090$). Instead of using the fall CMB power spectrum, which is computationally expensive, we often use a few compressed parameters (like $R,~l_a,~z_A)$ that capture most of the cosmological information, especially for background dynamics (e.g, $H_0$, $\Omega_m$, $\Omega_k$ etc.). The shift parameter $R$ which effectively shifts the angular scale of acoustic peaks in the CMB, encapsulating low matter content and geometry affect these positions. This is given as
\begin{equation}
    R =\sqrt{\Omega_m H_0^2} \frac{D_A (z_A)}{c}~~,
\end{equation}
where $D_A(z_A)$ is coming angular diameter distance to the last scattering surface, to be measured in a flat universe as 
\begin{equation}
    D_A (z) = \frac{1}{1+z} \int^z_0 \frac{c dz'}{H(z')}
\end{equation}
 using Planck 2018 compressed likelihood \cite{Planck2018} we obtain $R = 1.7492 \pm 0.0049$
CMB shift parameter is defined as 
\begin{equation}
    R= \Omega_m^{\frac{1}{2}} \int \limits_0^{z_{ls}} \frac{dz}{E(z)}~~~,
\end{equation}
where $z_{ls}$ is the redshift at last scattering.

The WMAP$7$ data gives us $R=1.726 \pm 0.018$ at $z=1091.3$ \cite{Eisenstein_2005, Planck2018, Wang_2013, Shafer_2015, Jimenez_2002, Wang_2007}. Chi squared in this case is defined as 
\begin{equation}
    \chi^2 _{CMB} = \frac{(R-0.469)^2}{(0.017)^2}~~~.
\end{equation}
For joint analysis, total chi squared function turns 
\begin{equation}
    \chi _{tot}^2 = \chi^2_{OHD}+\chi^2_{BAO}+ \chi^2_{CMB}~~~.
\end{equation}
Therefore, the best fit values of different DE model parameters are obtained from Table 1, as we have calculated this particular values in the last column of table 2. Also, different sigma confidence contours are given in fig 1a-1c, 2a-2c, 3a-3c, 4a-4c.
 

\begin{table}[h!]
    \centering
    \caption{Different redshift parametrization models with the EoS and best fit values.}
    \begin{tabular}{||>{\centering\arraybackslash}m{1cm}| >{\centering\arraybackslash}m{6.5cm} | >{\centering\arraybackslash}m{9.5cm} ||}
    \hline
    {\bf Sl. No.} & {\bf Model} & {\bf Best fit values of free parameters} \\[1ex]
    \hline
    1 & Linear Redshift parametrisation ({\bf LRP}) \cite{HutererTurner_2001} & $\chi^2_{CC}=4600.241321$, $\omega_0^{LRP}= -0.9985_{-0.1397}^{+0.1527}$, $\omega_1^{LRP}=-0.9664_{-0.3938}^{+0.4334}$~~~~; \\
    \cline{2-2}
    & $\omega^{LRP}(a) = \omega_0 ^{LRP} + \omega_1 ^{LRP}  \times \left({\frac{1-a}{a}}\right)$ & $\chi^2_{CC+BAO} = 5361.34598$, $\omega_0^{LRP}=-0.9934_{-0.1684}^{+0.1711}$, $\omega_1^{LRP}=-0.9995_{-0.5210}^{+0.5419}$~~~~; \\
    \cline{2-2}
    & $\omega^{LRP}(z) = \omega_0 ^{LRP} + \omega_1 ^{LRP} z$ & $\chi^2_{CC+BAO+CMB}=14556.024992$, $\omega_0^{LRP}=-0.9900_{-0.1516}^{+0.1542}$, $\omega_1^{LRP}=-0.9960_{-0.5038}^{+0.4784}$~~~~ ; \\
    \hline
    2 & Chevallier-Polarski-Linder Redshift parametrisation ({\bf CPL}) \cite{Chevallier2001, Linder_2003} & $\chi^2_{CC}=4606.814534$, $\omega_0^{CPL}=-0.9983_{-0.2647}^{+0.2593}$, $\omega_1^{CPL} = 0.9960_{-0.8338}^{+0.8336}$~~~~;\\
    \cline{2-2}
    & $\omega^{CPL}(a) = \omega_0^{CPL}+\omega_1^{CPL} \times  (1-a)$ & $\chi^2_{CC+BAO}=5367.92526$, $\omega_0^{CPL}= -0.9935_{-0.2083}^{+0.2034}$, $\omega_1^{CPL}= 0.9844_{-0.9425}^{+0.8762}$ ~~~~;\\
    \cline{2-2}
    & $\omega^{CPL}(z) = \omega_0^{CPL}+\omega_1^{CPL} \times \frac{z}{1+z}$ & $\chi^2_{CC+BAO+CMB}=14562.604273$, $\omega_0^{CPL}=-0.9930_{-0.2672}^{+0.2670}$, $\omega_1^{CPL} = 1.3007_{-1.2234}^{+1.2454}$ ~~~~;\\ 
    \hline
    3 & Jassal-Bagala -Padmanabhan Redshift Parameterization({\bf JBP}) \cite{JassalBaglaPadmanabhan_2005}&$\chi^2_{CC}=4611.217324$, $\omega_0^{JBP}=-0.9935_{-0.2854}^{+0.2951} $, $\omega_1^{JBP}=-0.9964_{-0.8803}^{+0.8897} $~~~~;\\
     \cline{2-2}
     & $\omega^{JBP}(a) = \omega_0^{JBP}+\omega_1^{JBP}\times a(1-a) $ & $\chi^2_{CC+BAO}=5372.32805$, $\omega_0^{JBP}=-0.9964_{-0.3882}^{+0.4225}$, $\omega_1^{JBP}=-1.4092_{-1.1413}^{+1.0850}$~~~~;\\ 
     \cline{2-2}
     & $\omega^{JBP}(z) = \omega_0^{JBP}+\omega_1^{JBP}\times \frac{z}{(1+z)^2}$ & $\chi^2_{CC+BAO+CMB}=14562.604273$, $\omega_0^{JBP}=-0.9930_{-0.2676}^{_0.2673},~\omega_1^{JBP}=-1.2124_{-1.0862}^{1.0046}$~~~~;\\
     \hline
     4 & Efstathiou Redshift Parametrization or, Log Parametrization({\bf ERP})\cite{Efstathiou_1999_EoS}& $\chi^2_{CC}= 4603.496460$, $\omega_0^{Log}=-0.9958_{-0.2008}^{+0.2055}$,
     $\omega_1^{Log}=0.9986_{-0.7090}^{+0.7365}$~~~~; \\
     \cline{2-2}
     & $\omega^{Log}(a) = \omega_0^{Log} -\omega_1^{Log} ln(a)$ & 
     $\chi^2_{CC+BAO}=5364.607187$, $\omega_0^{Log}=-0.9955_{-0.2295}^{+0.2382}$, $\omega_1^{Log}=-0.9977_{-0.7577}^{+0.7773}$~~~~;\\
     \cline{2-2}
     & $\omega^{Log}(z) = \omega_0^{Log}+\omega_1^{Log} ln(1+z)$ &$\chi^2_{CC+BAO+CMB}$, $=14556.513247$, $\omega_0^{Log}=-0.9989_{-0.2790}^{+0.2769}$, $\omega_1^{Log}=-1.0439_{-0.8637}^{+1.0063}$ \\
    \hline
\end{tabular}
\end{table}

\begin{figure}[h!]
    \centering
    \subfloat[For H(z)-z data Analysis] {\includegraphics[height=3.2in,width=3.2in]{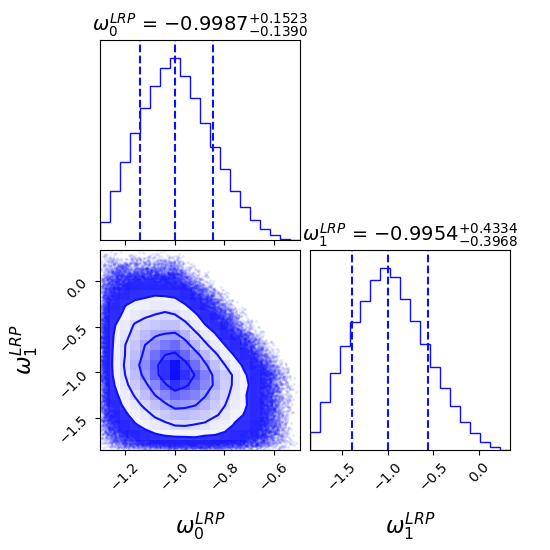}}\hspace{0.5cm}
   \subfloat[For H(z)-z+BAO Analysis] {\includegraphics[height=3.2in,width=3.2in]{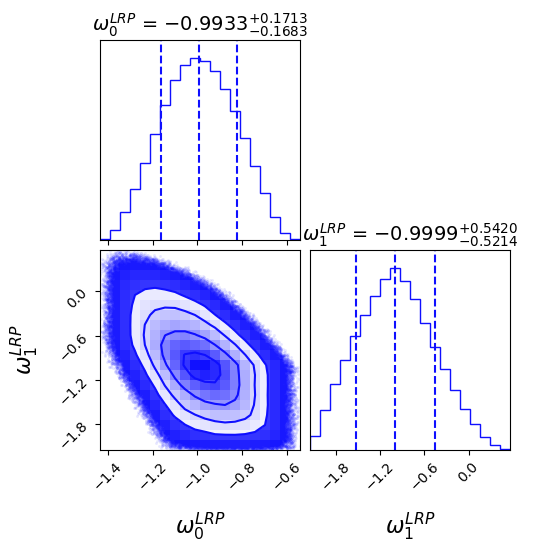}}  \\
    \subfloat[For H(z)-z+BAO+CMB Analysis] {\includegraphics[height=3.2in,width=3.2in]{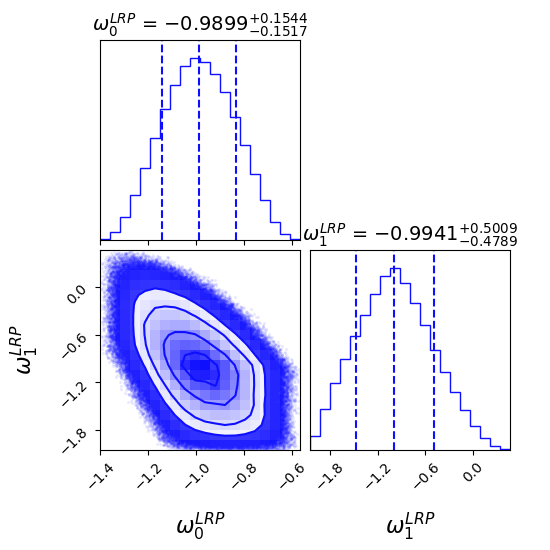}}\hspace{0.5cm}
    \subfloat[$\log(\frac{M}{M_0})$ vs redshift plot] {\includegraphics[height=3.2in,width=3.2in]{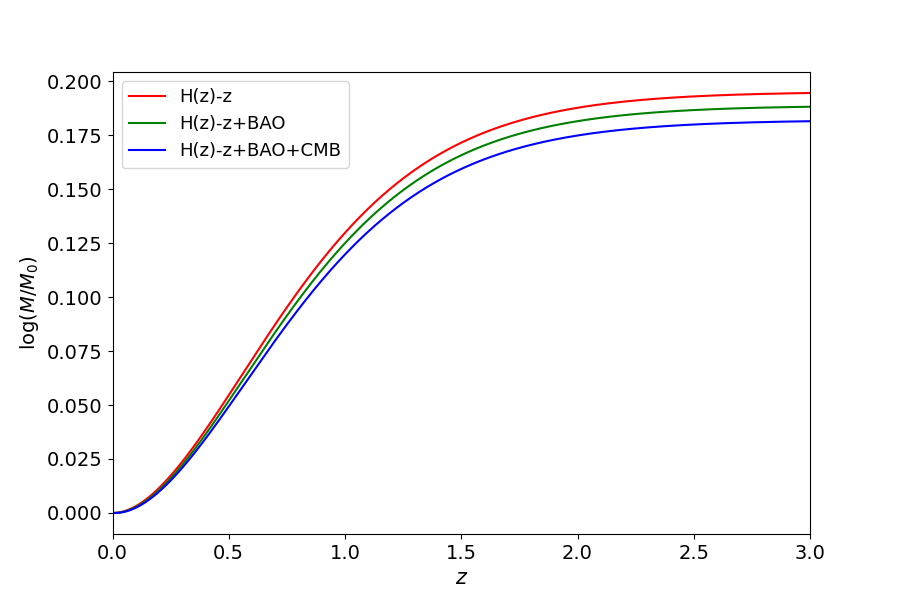}}
    \caption{1(a-c) represents confidence contours in $\omega_0^{LRP}-\omega_1^{LRP}$ plane and individual distributions of free parameters $\omega_0^{LRP}~\text{and}~\omega_1^{LRP}$. 1(d) represents the variation of the $\log\left\{\frac{M}{M_0}\right\}$ due to the accretion of LRP type DE onto a MCG contaminated BH.}
\end{figure}
In figure 1a to 1c, we have plotted the confidence contours of the free parameters and Gaussian distribution of the each of the parameters $\omega_0^{LRP}$
and $\omega_1^{LRP}$ for the model table 2-row 1 and data incorporated in table 1. In fig 1a, both the distributions of  $\omega_0^{LRP}$ and $\omega_1^{LRP}$ are positively/right skewed. The probability distributions of both parameters have a longer tail towards higher values. The parameter’s most likely (mode) value is lower than its mean. In parameter estimation, this suggests that while smaller values are preferred, the data still allows for a subset of much larger values with non-negligible probability. This supports presence of phantom to be likely. Fig 1b shows approximately  symmetric distribution of $\omega_0^{LRP}$. In fig 1c, $\omega_0^{LRP}$ is slightly left skewed. If the posterior distribution of the present-day DE EoS parameter $w_0$ is left-skewed, the most probable values correspond to $w_0 > -1$, representing a quintessence-like behavior of DE. However, the extended tail toward lower values ($w_0 < -1$) implies that a phantom-like regime is statistically allowed, albeit with lower probability. This asymmetry indicates that current observational data primarily favor a mildly dynamical DE model close to the quintessence domain, while still permitting limited phantom crossing within the uncertainty bounds.
 
\begin{figure}[h!]
    \centering
    \subfloat[For H(z)-z data Analysis] {\includegraphics[height=3.2in,width=3.2in]{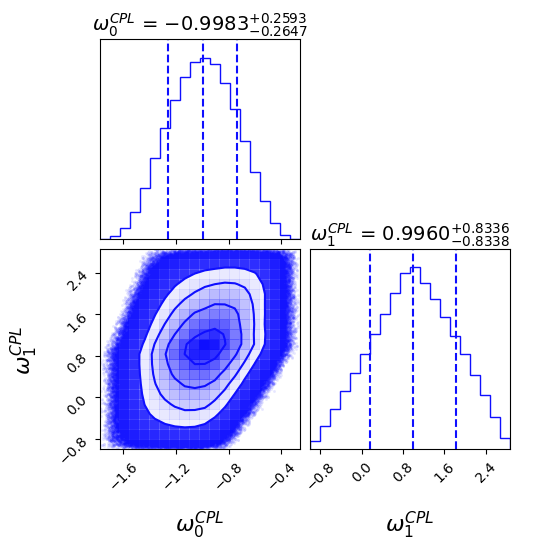}}\hspace{0.5cm}
   \subfloat[For H(z)-z+BAO Analysis] {\includegraphics[height=3.2in,width=3.2in]{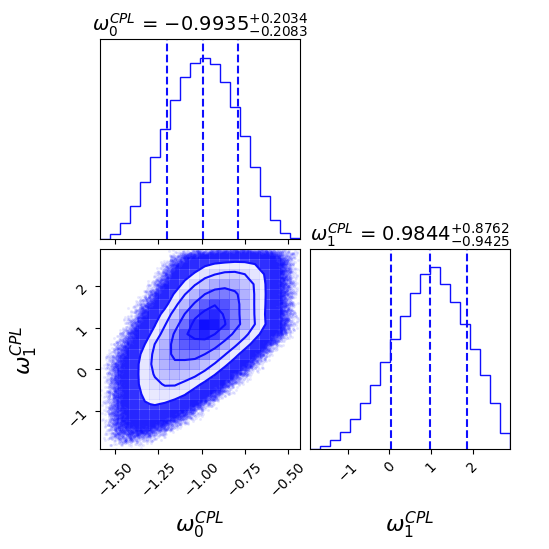}}  \\
    \subfloat[For H(z)-z+BAO+CMB Analysis] {\includegraphics[height=3.2in,width=3.2in]{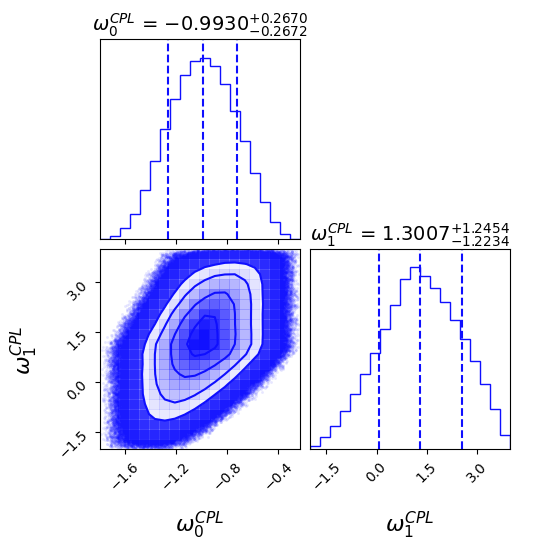}}\hspace{0.5cm}
    \subfloat[$\log(\frac{M}{M_0})$ vs redshift plot] {\includegraphics[height=3.2in,width=3.2in]{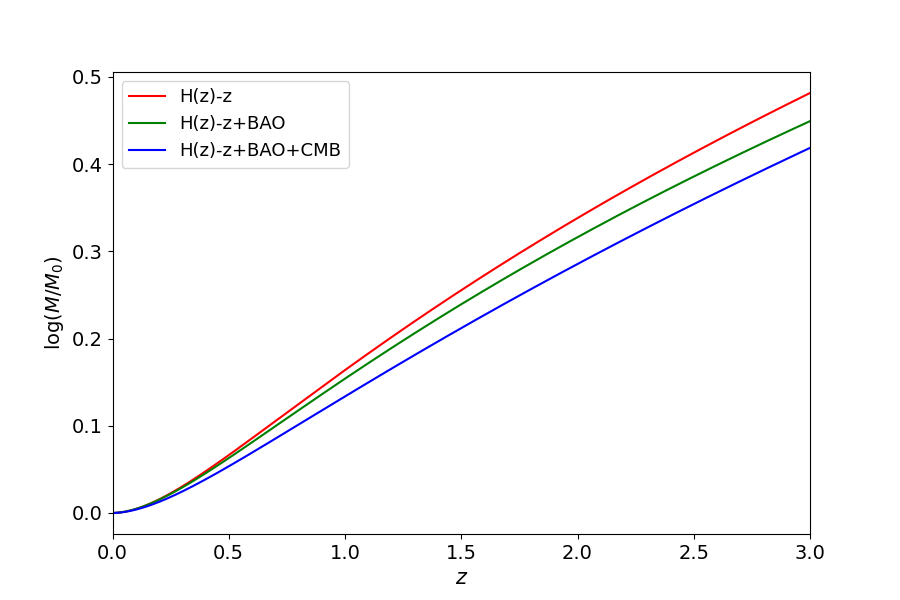}}
    \caption{2(a-c) represents confidence contours in $\omega_0^{CPL}-\omega_1^{CPL}$ plane and individual distributions of free parameters $\omega_0^{CPL}~\text{and}~\omega_1^{CPL}$. 2(d) represents the variation of the $\log\left\{\frac{M}{M_0}\right\}$ due to the accretion of CPL type DE onto a onto a MCG contaminated BH.}
\end{figure}
Figure 2a-c are free parameter contraining plots for CPL model. Here, both $\omega_0^{CPL}$ and $\omega_1^{CPL}$ are followed to be left skewed. The latter tilts more. A left-skewed posterior implies that larger values of the parameter are more probable, while there exists a non-negligible tail extending toward smaller values. In such a distribution, the most likely value (mode) lies above the mean, 
indicating that the lower parameter regime is less favored but remains statistically allowed. Cosmologically interpreting, phantom era is not likely with this model. Quintessence gets a good cheer up by the distribution of free parameters.

\begin{figure}[h!]
    \centering
    \subfloat[For H(z)-z data Analysis] {\includegraphics[height=3.2in,width=3.2in]{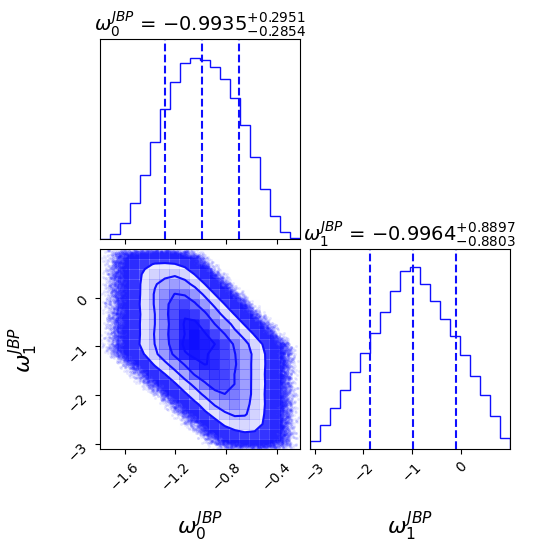}}\hspace{0.5cm}
   \subfloat[For H(z)-z+BAO Analysis] {\includegraphics[height=3.2in,width=3.2in]{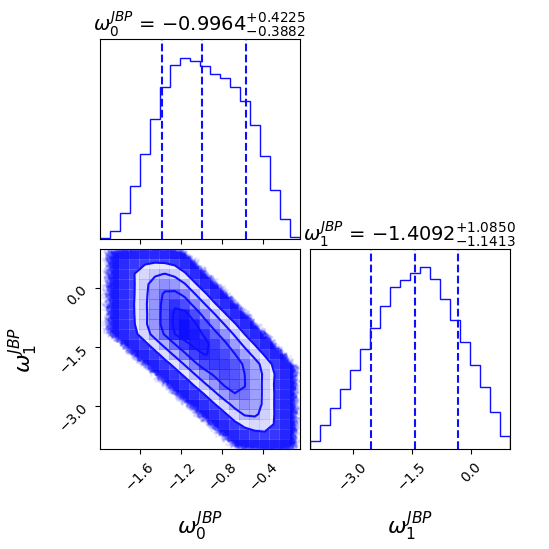}}  \\
    \subfloat[For H(z)-z+BAO+CMB Analysis] {\includegraphics[height=3.2in,width=3.2in]{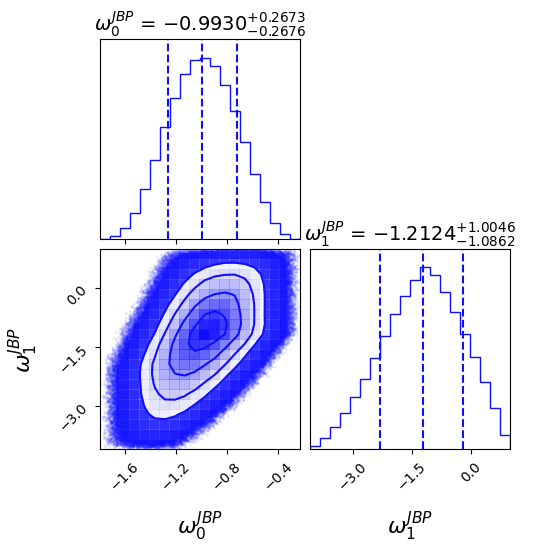}}\hspace{0.5cm}
    \subfloat[$\log(\frac{M}{M_0})$ vs redshift plot] {\includegraphics[height=3.2in,width=3.2in]{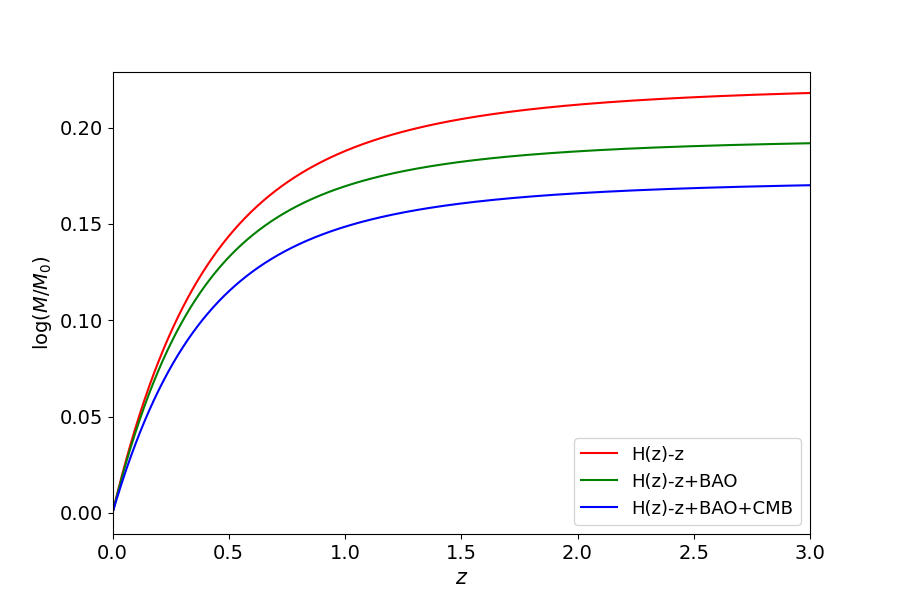}}
    \caption{3(a-c) represents confidence contours in $\omega_0^{JBP}-\omega_1^{JBP}$ plane and individual distributions of free parameters $\omega_0^{JBP}~\text{and}~\omega_1^{JBP}$. 3(d) represents the variation of the $\log\left\{\frac{M}{M_0}\right\}$ due to the accretion of JBP type DE onto a MCG contaminated BH.}
\end{figure}
In figure 3a-c, we have plotted different confidence contours of $\omega_0^{JBP}$ and $\omega_1^{JBP}$. Nature of free parameters' distribution match those of the CPL case. Only, $H(z)-z+BAO$ case evolves a convolution of two distributions. If the two component means differ significantly, then the two peaks remain distinct, producing a bimodal posterior. Here, two distinct physical regimes may become comparably probable. In a cosmological context, this situation can occur when the data admit 
two equally good fits - one corresponding to the $\Lambda$CDM limit 
($w_0 \approx -1$) and another lying in the phantom regime ($w_0 < -1$). 
Such a scenario signals a degeneracy in the parameter space, 
where the likelihood surface develops two nearly equivalent maxima.

\begin{figure}[h!]
    \centering
    \subfloat[For H(z)-z data Analysis] {\includegraphics[height=3.2in,width=3.2in]{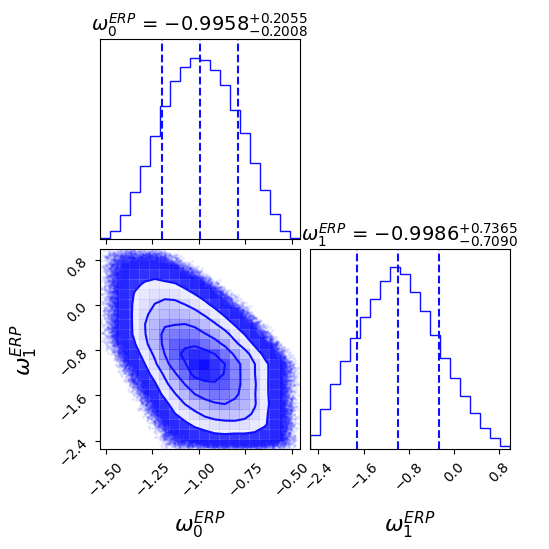}}\hspace{0.5cm}
   \subfloat[For H(z)-z+BAO Analysis] {\includegraphics[height=3.2in,width=3.2in]{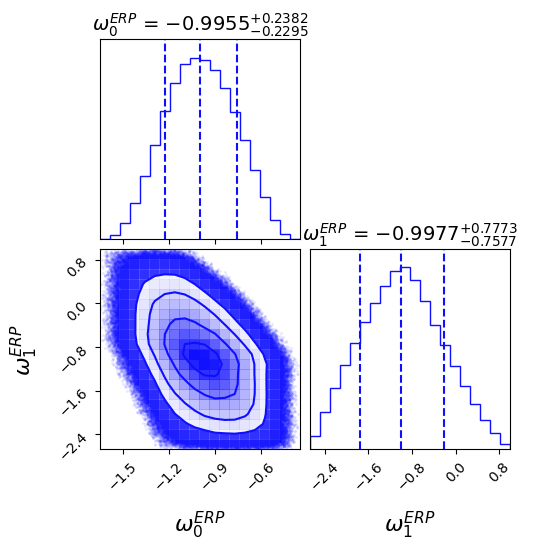}}  \\
    \subfloat[For H(z)-z+BAO+CMB Analysis] {\includegraphics[height=3.2in,width=3.2in]{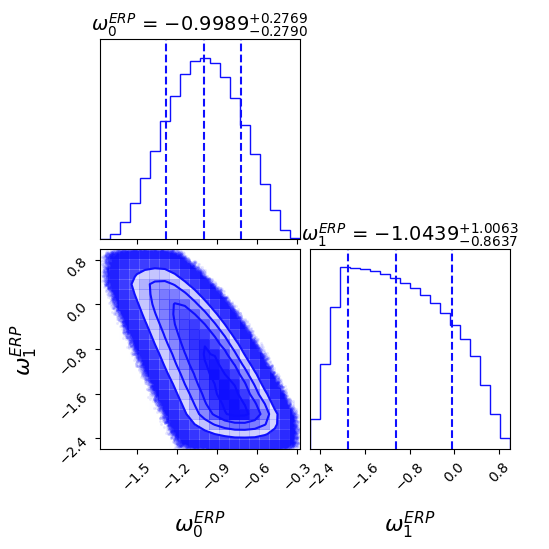}}\hspace{0.5cm}
    \subfloat[$\log(\frac{M}{M_0})$ vs redshift plot] {\includegraphics[height=3.2in,width=3.2in]{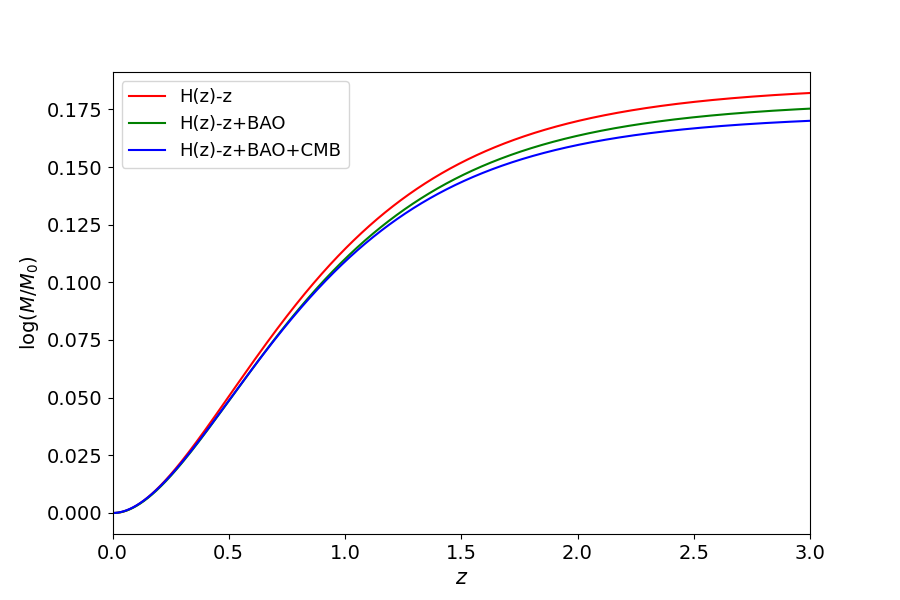}}
    \caption{4(a-c) represents confidence contours in $\omega_0^{ERP}-\omega_1^{ERP}$ plane and individual distributions of free parameters $\omega_0^{ERP}~\text{and}~\omega_1^{ERP}$. 4(d) represents the variation of the $\log\left\{\frac{M}{M_0}\right\}$ due to the accretion of ERP type DE onto a MCG contaminated BH.}
\end{figure}
In figure 4a-c, we will plot the Gaussian distributions and confidence contours of Efstathiou redshift parametrization model's free parameters. A joint analysis of Hubble data and BAO observations provides almost symmetric constraints on the free parameters. Only when CMB is added, the symmetric distribution of $\omega^{ERP}_1$ does not follow and turns extremely right skewed. An extremely right-skewed posterior distribution indicates that the most probable values of the parameter lie near the lower bound of its domain, while a long tail
extends toward higher values. In the context of dark energy models, this behavior 
implies that the data strongly favor a small or nearly vanishing deviation from the 
$\Lambda$CDM limit, yet allow the possibility of larger parameter values with 
progressively lower probability. Such asymmetry often arises from one-sided priors, 
nonlinear model responses, or the degeneracy of cosmological observables at large 
parameter magnitudes, reflecting a preference for minimal modification with an 
extended allowance for more exotic regimes.

\section{Dark Energy Accretion towards Black Holes}

Let $u^{\mu} \equiv \dfrac{d x^\mu}{d \tau}$ be the four speed for the $\mu$-th coordinate $x^\mu$ and $\tau$ is the proper time. Using the dot product property of four vectors $u_\mu u^\mu =1$. In addition, DE accretion is supposed not to shed any influence on the corresponding BH's spherical symmetry. This makes $u^\theta$ and $u^\phi$ to vanish. For simplicity, $u^r =u<0$ (due to infall) is chosen and hence the non zero velocity terms can be written as 
\begin{equation}
    u_r= g_{r r} u^r =\frac{u}{f_{MCG}(r)}~,~~u^t = \frac{\sqrt{f_{MCG}(r)+u^2}}{f_{MCG}(r)}=\sqrt{\frac{1}{f_{MCG}(r)}+\frac{u^2}{\{f_{MCG}(r)\}^2}}~~~\text{and}~~u_t = - \sqrt{f_{MCG}(r)+u^2}~~.
\end{equation}

With these, we construct the components of the stress energy tensor as.
\begin{equation}
\left.{
\begin{split}
    T_t^t &= (\rho +p) u^t u_t +p =-\rho - \frac{u^2}{f_{MCG}(r)}(\rho +p)~~,\\
    T_r^r &= (\rho+p) u^r u_r +p = (\rho + p) \frac{u^2}{f_{MCG}(r)}+p~~,\\
    T_\theta^\theta &= T_\phi^\phi =p ~~\text{and}~~T_t^r = - u(\rho + p) \sqrt{f_{MCG}(r) +u^2}~~~.
    \end{split}~~~~~~~~~~~}\right\rbrace
\end{equation}
Along with the 4-current conservation of mass flux, $\dfrac{\partial J^\mu}{\partial x^\mu}=0$ will be simplified now.
\begin{equation}
    J^\mu = \rho \dfrac{dx^\mu}{d \tau} = \rho \begin{pmatrix}
        c \\
        \dfrac{d x^i}{d \tau} 
    \end{pmatrix} = \begin{pmatrix}
        c \\
        u\\
        0\\
        0
    \end{pmatrix} ~~\Rightarrow~~\rho u r^2 = \xi_0~,~~\text{a constant}~~\mu = t,~r,~\theta,~\phi~;~~i=r,~\theta,~\phi~.
\end{equation}

Now we will only opt the $t$-th component of $T_{\mu;\nu}=0$ (`;' signifies covariant differentiation) i.e., $T_{t;\nu}^\nu =0$, and we obtain
\begin{equation} \label{0th equation}
    0= T_{t;\nu}^\nu = T^t_{t;t}+ T^r_{t;r}+T^\theta_{t;\theta} + T_{t;\phi}^\phi
\end{equation}
\begin{align}
\Rightarrow~~& ur^2 (\rho +p) \sqrt{f_{MCG}(r)+u^2} M^{-2} =\xi_1~,~~~\text{another constant of integration~~.}\\ \label{modified 0th component}
 \end{align}

We require another equation to constrain our three parameters $u$, $r$ and $\rho$. For this, we will take the projection of conservation equation on four velocity vectors, i.e., $u^\mu T_{\mu ; \nu}^\nu =0$ and obtain \cite{Debnath_2015} as
\begin{equation} \label{0th equation in x}
    ur^2 M^{-2} \exp{\left\lbrace{\int_{\rho_\infty}^{\rho} \dfrac{d \rho^\prime}{\rho^\prime +p(\rho^\prime)}}\right\rbrace}=-\xi_2 = ux^2 \exp{\left\lbrace{\int_{\rho_\infty}^{\rho} \dfrac{d \rho^\prime}{\rho^\prime +p(\rho^\prime)}}\right\rbrace}~~~~,
\end{equation}
$\xi_2>0$ symbolizes a constant related to the energy flux \cite{Debnath_2015,Babichev_2004, Babichev_2005, Babichev_2013}. Here $\rho$ and $\rho_\infty$ respectively represent the values of the density at a finite distance and at an infinite distance. $x$ is a dimensionless radial distance parameter. 

Suppose the number density of the medium is $n$ \cite{Babichev_2005} then,
\begin{equation}
    \dfrac{d \rho}{\rho+p} =\dfrac{d n}{n}~~.
\end{equation}
Integrating, \begin{equation} \label{n ratio}
    n\equiv n_\infty\exp{\left\lbrace{\int_{\rho_\infty}^{\rho} \dfrac{d \rho^\prime}{\rho^\prime +p(\rho^\prime)}}\right\rbrace}~~~,
\end{equation}
where $n_\infty$ is the concentration of DE \cite{Babichev_2005} \eqref{0th equation}  and \eqref{modified 0th component} evolves
\begin{equation} \label{equation of lambda3}
    \frac{\rho +p}{n} \sqrt{u^2+ f(r)} =\xi_3 =-\frac{\xi_1}{\xi_2} =\frac{\rho_\infty +p(\rho_\infty)}{n_\infty}~~~.
\end{equation}

To determine $\xi_2$ which has the dimension of energy flux, effective sonic speed squared is evaluated as 
\begin{equation}
    v^2 = \frac{n}{\rho+p} \dfrac{d(\rho+p)}{dn}-1~~~.
\end{equation}

Differentiating equations \eqref{0th equation in x} and \eqref{equation of lambda3} and using \eqref{n ratio}, we obtain
\begin{equation}
    \dfrac{du}{u} \left[{v^2 -\frac{u^2}{u^2+f(r)}}\right] + \dfrac{dx}{x} \left[{2 v^2 -\frac{x f^\prime (r)}{2 (f(r)+u)}}\right]=0~~~.
\end{equation}

From this equation, radial inward speed gradient can be calculated as a numerator to denominator ratio. The denominator will clearly vanish at some $x=x_c$ in the interval $(0, ~\infty)$ depending on the structure of $f(r)$. However, to make the flow physical, numerator should vanish parallely. These two vanishing expressions give us the relations
\begin{equation} \label{u_c equation}
    u_c^2 = \frac{1}{4}x_c \dfrac{d f}{dx} \Bigm|_{x=x_c}~~\text{and}
\end{equation}
\begin{equation}\label{v_c equation}
    v_c^2= \frac{x_c }{x_c f^\prime (x_c)+ 4 f(x_c)} \left. \dfrac{d f^\prime}{dx}\right\vert_{x=x_c} ~~~.
\end{equation}

Now, using equations \eqref{u_c equation} and \eqref{v_c equation}, from \eqref{modified 0th component} we have
\begin{equation}
    \frac{\rho_c+p(\rho_c)}{\rho_\infty +p(\rho_\infty)} =\frac{\exp{\left\lbrace{{\Huge\int}_{\rho_\infty}^{\rho} \dfrac{d \rho^\prime}{\rho^\prime +p(\rho^\prime)}}\right\rbrace}}{2 \sqrt{x_c \left. \dfrac{\partial f(x)}{\partial x}\right\vert_{x=x_c}+ 4 f(x_c)}}~~~.
\end{equation}
We are able to calculate 
\begin{equation}
    \xi_2 = \left. {\frac{x_c^3}{2} \dfrac{df}{dx}}\right\vert_{x=x_c} \times \exp{\left\lbrace{\int_{\rho_\infty}^{\rho_c} \dfrac{d \rho^\prime}{\rho^\prime +p(\rho^\prime)}}\right\rbrace}~~~.
\end{equation}
\begin{equation}
    or,~~\xi_2 = \left. {\frac{x_c^3}{2} \dfrac{df}{dx}}\right\vert_{x=x_c} \times \exp{\left\lbrace{\int_{\rho_\infty}^{\rho_c} \dfrac{d \rho^\prime}{\rho^\prime +\left({\omega_0+\omega_1 z}\right) \rho^\prime}}\right\rbrace}~~~.
\end{equation}

For different DE models, different values of $\xi_2 \left({=\xi_2^{DE}~say}\right)$ can be obtained.

Using this, the rate of change of mass of the BH is calculated as 
\begin{equation}
    \dot{M} =-4 \pi \xi_1 M^2 = 4 \pi \xi_2 M^2 \left[{\rho_\infty+p(\rho_\infty)}\right]~~.
\end{equation}

Several articles \cite{Abbas_2013, Babichev_2008, Jamil_2010, Jamil_2010_GRG, Dutta_2019} suggest that the rate of change of mass can be calculated does not satisfy the dominant energy condition, whenever we can apply the EoS from equation (\ref{DEEoS}).

Hence, as the proportionality constants are independent of the distance from the central gravitating engine, we generalise the result to \begin{equation}\label{FR}
    \dot{M} = 4 \pi \xi_2 M^2 [\rho+p(\rho)]~~~.
\end{equation}

Sign of $\dot{M}$ is solely determined by that of $(\rho+p)$. For quintessence $-1<\frac{p}{\rho}<-\frac{1}{3}$ and hence $\dot{M}>0$. But once phantom barrier $\left({\frac{p}{\rho}=-1}\right)$ is crossed, $p+\rho<0$ and hence $\dot{M}<0$. Using equation (\ref{FR}) and equation of continuity, we obtain \cite{Basak:2025ysq}
\begin{align*}
    \dfrac{d M}{d \rho} =-\frac{4 \pi \xi_2 M^2}{3H^2} ~~\Rightarrow~~&\int\limits_{M}^{M_0} \dfrac{dM}{M^2} = - \frac{4 \pi \xi_2}{3} \int\limits_{\rho}^{\rho_0} \dfrac{d \rho}{H} 
    \Rightarrow~~ \frac{1}{M} -\frac{1}{M_0} = - \frac{4 \pi \xi_2}{3} \int\limits_{\rho}^{\rho_0} \dfrac{d \rho}{H}
\end{align*}
\begin{equation}
    \Rightarrow~~ M =\frac{M_0}{1+ \frac{4 \pi \xi_2 M_0}{3} \int\limits_{\rho}^{\rho_0} \frac{d \rho }{\sqrt{ \frac{8 \pi G_N}{3} \rho+ \frac{\Lambda}{3}- \frac{\kappa}{a^2}}}}~~~.
\end{equation}

Here $M_0 \left(M(z=0)\right)$ is the present day mass of the BH. The ernergy density of present time $\rho(z=0)$ is constituted of three parts : the matter density of present time $\rho_{m_0}$, radiation density $\rho_{rad_0}$ and the same for DE $\rho_{DE_0}$.

Hence we obtain 
\begin{equation}
    M = \frac{M_0}{1+ \frac{4 \pi \xi_2 M_0}{3} \int\limits_{\rho_0}^\rho \frac{d \left(\rho_b + \rho_{DE}+ \rho_r\right)}{\sqrt{ \frac{8 \pi G_N}{3} \left(\rho_b + \rho_{DE}+ \rho_r\right)+ \frac{\Lambda}{3}- \frac{\kappa}{a^2}}}}~~.
\end{equation}

In figure 1d, $\log\left(\frac{M}{M_0}\right)$ vs $z$ is plotted for LRP model$(\rho^{LRP}{(z)}  = \rho_{\phi0}~exp(3 \omega_1 z)(1 + z)^{3(1 + \omega_0 + \omega_1)})$. In past, i.e., for high redshift, almost 20\% excess mass than today is observed. Mass started to get reduced as we shift towards the zero redshift. In a LRP, the observed decrease of 
$\log_{10}[M(z)/M_{0}]$ from $0.2$ to $0$ as $z$ evolves from $2$ to $0$ 
suggests that the relative mass growth of BHs diminishes toward the 
present epoch, reflecting a gradual suppression of accretion efficiency. When 
the BH resides in a MCG background, this trend 
can be understood as a dynamical consequence of the evolving equation of state 
of the MCG: at higher redshifts ($z\sim2$), the MCG behaves like a matter like 
fluid, enabling efficient accretion, whereas at lower redshifts the transition 
to a negative pressure, DE like phase produces a repulsive effect that 
hinders infall. Consequently, the decline in $\log_{10}[M(z)/M_{0}]$ captures 
the cosmological shift from a dense, accretion dominated era to a DE 
dominated regime, consistent with the cosmic downsizing scenario of BH 
growth.

In figure 2d, $\log\left(\frac{M}{M_0}\right)$ is plotted with respect to $z$ and we observe a steep fall towards origin. For the CPL DE parametrization $\rho^{CPL}_{\phi} (z) = \rho^{CPL}_{\phi 0}~ \exp\bigg\{-3w_1^{CPL}~\frac{z}{1+z}\bigg\}(1+z)^{3\left(1+w^{CPL}_0+w^{CPL}_1\right)}$, a steep 
decline of BH mass in the past during accretion onto a MCG contaminated background indicates that the effective energy density of 
the surrounding fluid was dominated by a strongly negative pressure component. 
At high redshifts, the CPL form $\omega(z) = \omega_{0} + \omega_{1}\,z/(1+z)$ 
allows $\omega(z) < -1$, leading to a phantom like regime where the inflowing MCG 
exerts repulsive gravitational effects, causing a rapid decrease in BH 
mass. This behaviour reflects the nontrivial coupling between the dynamical DE sector and accreting matter, implying that the BH evolution is 
sensitive to both the background equation of state and its redshift evolution. Physically, this represents the extraction of 
gravitational energy by a fluid with supernegative pressure, violating the usual 
mass growth trend expected in standard accretion. Such a phenomenon indicates that 
in a DE dominated background, the thermodynamic balance between the BH and the cosmic fluid is altered, with the horizon effectively acting as an 
energy emitter rather than an absorber.

In figure 3d, log of BH mass to present time BH mass ratio is plotted as a function of redshift. For Linear and CPL, near origin slopes were less. For JBP $\left(\rho^{JBP}_{\phi}(z) = \rho^{JBP}_{\phi 0}~ \exp\bigg\{\frac{3w^{JBP}_1}{2(1+z)}\bigg\}(1+z)^{3(1+w^{JBP}_0)}\right)$ near origin slope is high. In an MCG dominated universe, the comparatively smaller near origin slopes for the 
Linear and CPL parametrizations imply that the DE equation of state evolves 
slowly at late times, producing a quasi-stationary background with weak pressure 
gradients around the BH. As a result, the inflow and outflow of energy across 
the horizon nearly balance, leading to a gentle and stable evolution of BH 
mass. However, the JBP parametrization, exhibiting a steep 
slope near the horizon, corresponds to a rapidly varying $\omega(z)$ at low redshift. 
This induces a stronger local pressure anisotropy in the MCG fluid, enhancing the 
energy flux and triggering a sharp change in the accretion dynamics. Physically, it 
suggests that in the JBP framework, DE interacts more actively with the 
BH environment, and the transition from matter-like to  DE dominated 
behaviour produces a more violent mass evolution near the horizon.

For ERP $\left(\rho^{ERP}(z)= \rho_{\phi0}~(1+z)^{3\left\{1+\omega_0^{Log}+\frac{\omega_1^{Log}}{2}\log(1+z)\right\}}\right)$, we follow the mass of the BH to fall with time. The pattern does match with that of LRP model. The resemblance between the $\log_{10}[M(z)/M_{0}]$ behaviour in the logarithmic and 
linear DE parametrizations arises from the fact that both models describe a 
gradual, smooth evolution of the equation of state $\omega(z)$ without introducing any 
sharp dynamical transitions in the cosmic background. In an MCG dominated universe, 
the accretion rate onto the BH is primarily governed by the pressure to density 
ratio of the surrounding fluid; hence, when $\omega(z)$ evolves linearly or logarithmically 
with redshift, the resulting pressure gradient and energy flux toward the horizon remain 
comparably weak and slowly varying. This leads to a nearly identical mass evolution profile 
because the effective gravitational potential around the BH changes adiabatically, 
maintaining thermodynamic equilibrium with the expanding MCG background. Physically, this 
implies that both parametrizations represent a cosmological regime where DE acts 
as a quasi-stationary, weakly dynamical field sufficient to drive acceleration but not strong 
enough to significantly disrupt the steady state accretion process of the BH.

\section{Conclusion}
{\bf Black Hole Mass Evolution in Dark Energy Backgrounds: Summary of Parametrization Effects :  }The evolution of black hole mass in the presence of a cosmic fluid, particularly 
a MCG, is highly sensitive to the choice of DE EoS parametrization. Observables such as the logarithmic mass 
ratio, $\log_{10}[M(z)/M_{0}]$, provide insights into the interplay between local 
horizon physics and global cosmological expansion. Across different parametrizations, 
distinct trends emerge, reflecting how the evolving pressure and energy density of 
the cosmic fluid regulate accretion dynamics.

{\bf Linear, Logarithmic, and CPL Parametrizations : }In linear ($\omega(z) = \omega_0 + \omega_1 z$) and logarithmic parametrizations, 
the mass ratio evolves smoothly with redshift. Both exhibit gentle slopes near the 
present epoch ($z \sim 0$), indicating that the DE EoS changes slowly, 
producing a quasi-stationary background where the inflow of energy toward the 
black hole remains moderate. The logarithmic parametrization, while slightly 
altering the functional form, produces nearly identical trends to the linear model, 
demonstrating that the accretion is governed primarily by the smooth, adiabatic 
variation of the MCG pressure. A small offset in amplitude (e.g., $\sim0.1$ order 
less) can arise due to minor differences in effective energy flux, but the qualitative 
behaviour remains unchanged. The CPL parametrization ($\omega(z) = \omega_0 + 
\omega_1 z/(1+z)$) also shows slow mass evolution near the origin, reflecting 
weak late-time pressure gradients that maintain a nearly steady accretion flow.

{\bf JBP Parametrization  : }The Jassal--Bagla--Padmanabhan (JBP) model, in contrast, exhibits a steep slope 
near the horizon at low redshift, indicating a rapid change in the EoS $\omega(z)$ 
as the universe evolves. This produces a stronger local pressure anisotropy in the 
MCG fluid, enhancing energy flux and causing a more pronounced variation in the 
black hole mass. Physically, it reflects a regime where DE interacts 
more actively with the black hole environment, amplifying the sensitivity of accretion 
to the transition between matter-like and dark energy-dominated phases.

{\bf Physical Interpretation Across Models : }Across all parametrizations, key physical insights emerge:

\begin{itemize}
\item The slope and amplitude of $\log_{10}[M(z)/M_{0}]$ encode the local pressure 
      gradients and energy flux at the black hole horizon, which are controlled by 
      the evolving EoS $\omega(z)$ and the MCG properties.
\item Models with gentle slopes near the origin (Linear, Log, CPL) reflect adiabatic, 
      quasi-static cosmic backgrounds, where dark energy evolves slowly and 
      mass growth proceeds steadily.
\item Steep slopes (JBP) signify rapid late-time evolution of the EoS, producing 
      stronger local energy fluxes and more sensitive black hole growth.
\item Minor amplitude shifts (e.g., 0.1 order) indicate changes in the effective 
      accretion efficiency or flux without altering the qualitative trend.
\item Overall, black hole mass evolution serves as a sensitive probe of the 
      interplay between local strong gravity and global cosmic expansion, capturing 
      the cumulative effects of accretion, mergers, and dark energy dynamics.
\end{itemize}

{\bf In a nutshell, } The comparative study of dark energy parametrizations—Linear, CPL JBP and Logarithmic within an MCG background reveals that both the slope and 
non-monotonic features of $\log_{10}[M(z)/M_{0}]$ carry deep physical meaning. 
Steep slopes indicate strong late-time dynamical effects, non-monotonic minima or 
maxima trace transitions in the repulsive gravitational influence of DE, 
and gentle slopes correspond to quasi-static accretion. Collectively, these results 
demonstrate that the evolution of black hole mass is a sensitive indicator of the 
time-dependent equation of state, the pressure-density interplay of the cosmic fluid, 
and the thermodynamic behaviour of black hole horizons in a universe dominated by both DM like and DE like components.



\section*{Acknowledgment} RB and FR thank Inter University Centre for Astronomy and Astrophysics (IUCAA), Pune, India, for granting Visiting Associateship. PB and SP thank the Department of Mathematics, Jadavpur University. And SD thanks the Department of Mathematics, The University of Burdwan, for different research facilities.

\section*{Data Availability Statement}
Data sharing not applicable to this article as no datasets were generated or analysed during the current study.

\section*{Conflict of Interest}

There are no conflicts of interest.

\section*{Funding Statement}

There is no funding to report for this article.

\section*{Code/Software}

No software/Coder was used in this study.



\bibliographystyle{ieeetr} 

\bibliography{references}

\end{document}